# Measuring the Credibility of Student Attendance Data in Higher Education for Data Mining

Mohammed Alsuwaiket, Christian Dawson, and Firat Batmaz

*Abstract*—Educational Data Mining (EDM) is a developing discipline, concerned with expanding the classical Data Mining (DM) methods and developing new methods for discovering the data that originate from educational systems.

Student attendance in higher education has always been dealt with in a classical way, i.e. educators rely on counting the occurrence of attendance or absence building their knowledge about students as well as modules based on this count. This method is neither credible nor does it necessarily provide a real indication of a student's performance.

This study tries to formulate the extracted knowledge in a way that guarantees achieving accurate and credible results. Student attendance data, gathered from the educational system, were first cleaned in order to remove any randomness and noise, then various attributes were studied so as to highlight the most significant ones that affect the real attendance of students. The next step was to derive an equation that measures the Student Attendance's Credibility (SAC) considering the attributes chosen in the previous step. The reliability of the newly developed measure was then evaluated in order to examine its consistency. Finally, the J48 DM classification technique was utilized in order to classify modules based on the strength of their SAC values.

Results of this study were promising, and credibility values achieved using the newly derived formula gave accurate, credible, and real indicators of student attendance, as well as accurate classification of modules based on the credibility of student attendance on those modules.

## I. INTRODUCTION

Student attendance in higher education is an important issue as it has been shown to directly affect students' performance [1], In order to consider student attendance as a criterion for estimating students' performance, student attendance data should be represented by a form that reflects the real weight of student attendance rather than the number of attendance occurrences, because of the fact that previous studies that considered investigating students' attendance have either considered the number of attendance occurrences as it is, or used its average value without relating these values to any other factors that may represent the real student attendance. On the other hand, those studies intended to use student attendance data without preparation that can neither be reliable nor credible. Therefore, the Student Attendance's Credibility (SAC) should be taken into consideration.

The credibility in research involves establishing that the results of research are credible or believable. It is a term that when incorporated with other terms like: Transferability, Dependability, and Confirmability, can be used to replace 'reliability' and 'validity', which are usually linked to quantitative research [2]. On one hand, the reliability refers to the ability to measure the attributes of a variable or construct consistency; while on the other hand, the validity is the extent to which the measurement of a concept is done accurately [3]. SAC represents a measure that can determine how much credible student attendance is for each class during the semester.

Hence, the ability to measure SAC is crucial in educational environments since it can accurately evaluate the weight of student attendance compared to his/her number of attendance occurrences. By measuring SAC for each module and comparing it to the same student's SAC of other modules, the performance of the student can then be evaluated and compared to his/her performance in other semesters. SAC is based on various factors that will be described in detail later in this paper.

By measuring SAC for each module he/she attends, modules can be classified using data mining classification techniques depending on SAC values tied to the module.

Data mining is exploration and analysis, by automatic or semi-automatic means, of large quantities of data in order to discover meaningful patterns [4]. Fig. 1 shows the process of data mining.

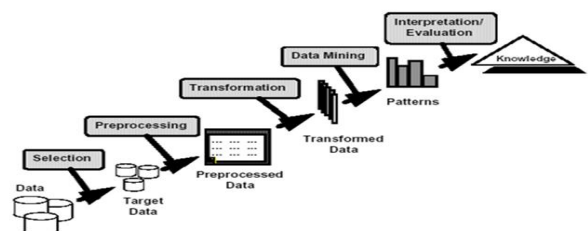

Fig. 1. Levels of the data processes [5].

Figure 1 shows that in order to start mining the data, one should select the target data from the database. Then, preprocess the target data to get the transformed data. Finally, mine the data to get the proposed knowledge.

In terms of data preprocessing, the initial attendance data used in this study were random, noisy and needed to be cleaned. If the data were not processed and unchanged, it would have been impossible to achieve logical results since the technique to be applied would never comprehend such data.

Based on that, an extensive research was undertaken on this field in order to investigate the existence of any related works that considered cleaning the attendance and absence data. The majority of those research works either neglected these data or used the data without cleaning or modification.

There are increasing research interests in using data mining techniques in higher education (HE), allowing to the EDM to emerge. EDM is concerned with developing different methods that discover knowledge from data from educational environments [6]. HE data can be collected from historical and operational data in the databases of educational institutes. The students' data can be personal or academic data. They also can be collected from e-learning systems which have a vast amount of information used by most institutes [7], [8].

The objectives of this paper, framed to assist the academic achievers in HE, are as follows:
1) Assess the importance of student attendance data in HE by creating a formula that weighs up the credibility of student attendance rather than using the number of attendance occurrences in evaluating the performance of students.
2) Identify the different factors that can affect SAC.
3) Construct a prediction model using datamining classification techniques in order to predict student attendance credibility for different modules and other future semesters.

## II. RELATED WORK

There is an increasing interest in using data mining in HE. Most of these research works are concerned with the performance of students [4] [9]. There are many resources for data in HE such as: graduate students' surveys, students' transcripts, workshops, etc. However, this paper concentrates initially on student attendance data since many previous studies have considered it as a significant indicator to student performance. However, it appears to be that no studies have taken the credibility of student attendance into consideration.

Many researchers have considered classifying student attendance data using Decision Tree (DT) data mining classification techniques [10]. By adopting student attendance rates as their criterion, they concluded that a lower attendance rate negatively affects the student's GPA (Grade Point Average). In addition, [11] used DM prediction techniques (classification, Predictive model, and Bayesian classification) to find the effective factors that determine a student's test grade, and then adjusted these factors to improve the performance of students' test grade.

Ref. [12] used decision tree classification on a student e-learning database to predict the student's division. The data set used in this study was obtained from students of an engineering college from 2006 to 2010 and contained data on 30 students. The main objective in this study was to help the students and the teachers to improve the grades of the students and identify those students which needed special attention to reduce failure percentages and take appropriate action for the upcoming evaluations.

After reviewing the related literature, it could be concluded that very few researchers were interested in the credibility of student attendance data, instead they emphasize the student attendance data as a number of occurrences. For example, the data collected for the Computer Science Department at Loughborough University in the UK during the academic year 2014-2015 cannot be considered usable because – for some modules -student attendances were taken only twice during a period of a whole semester (11 weeks). In such cases, if a student attended the module when the attendance was taken, his/her attendance average would be 100% even if he/she was not attending the module other than those two times. This cannot give a real indication of student performance.

## III. STUDENT ATTENDANCE'S CREDIBILITY (SAC)

Data mining needs to have clean and reliable data to achieve acceptable results. After preparing the student attendance data, it then has to be sorted and classified. When dealing with student attendance data, data were found to be noisy and in need of cleaning. In contrast to other research works which dealt with student attendance data without any preparation, this study considers preparation of the data as a first step before applying any data mining technique.

After preparing the data, the next step was to investigate which parameters may play a role in formulating the credibility of student attendance. The result of studying the most effective parameters is a proposed formula that can be applied to student attendance data to achieve credible results which in turn may help in evaluating the modules in terms of student attendance by measuring the credibility of each module and compare the results.

Measuring SAC in HE can give an indication on the credibility of students' attendance, and can help in:
1) Achieving credible data source for difference data mining techniques.
2) Evaluating the modules by accurately determining the student attendance and absence not by their number of occurrences, but rather depending on the effective weights of attendance and absence.
3) Evaluating the credible student attendance for each module and comparing it to the same student's records in different years.
4) Evaluating the performance of students based on their credible attendance or absence in each module, and then for all years of study.

There are many forms of students' attendance that are used in the higher education institutes such as: office hours, supervisors meetings, workshops, etc. However, this study concentrates on student attendance in modules (classes).

The main objective of this work is to measure the credibility of each module depending on two main factors: the average of student attendance and the number of attendance occurrences taken by the instructor during the semester.

A credibility equation has been formulated in order to measure the credibility of student attendance; this equation will be described step by step in the following section.

### A. SAC Equation

This section describes and validates the equation of SAC. This equation helps us to measure the credibility of student attendance; it can tell whether the module has a good attendance rate for a specific semester or not depending on the following parameters:
1) Average of student attendance for each module: it is a significant parameter that represents the ratio between

the number of students attended at a certain week in each module to the number of registered students in each module. Hence it is a variable parameter that relies on both numbers.

2) Number of attendance occurrences taken by the instructor for each module per semester: this significant parameter represents the number of attendance occurrences taken by instructor, and its significance comes from the fact that no information of the student attendance will be available whether the students really attended the module or not.

3) Total number of weeks per semester: a variable parameter, depending on the system of each school, and it should be part of the equation since it is connected directly to both the number of attendance occurrences taken by instructor, and the average of student attendance for each module.

*1) Average of student attendance*

As mentioned earlier, the average of student attendance depends on two factors: number of students attended at certain week and the number of registered students at each module. Number of students registered on a module is given, hence let us assume that:

X= Number of students attended on a certain week at a module
Y= Number of registered students at a module
- Number of weeks starting with 1 until the last week (n)

By dividing X over Y, we can obtain the ratio of attendance for one week only. Therefore, by considering the summation of this value throughout the semester multiplied by 100%, we can calculate the average percentage of student attendance for the whole semester.

To find the average of student attendance, we have to use the following equation:

$$\text{Average of Student Attendance} = \sum_{week=1}^{n} \left(\frac{X}{Y}\right) * 100\% \quad (1)$$

Note: the number of registered students in a module must be more than zero (otherwise there would be no module), therefore:

$$Y \neq 0$$

Note: Number of students attended at a certain week in such module must be less or equal to the number of registered students in such module, therefore:

$$X \leq Y$$

Average ∈ [0-100%]

We conclude that the more students attend at a certain week will give a better average of attendance (closer to 100%).

*2) Credibility of student attendance*

In order to measure SAC for each module, three parameters should be considered: the average value of student attendance should be calculated for each module; the number of attendance occurrences taken by the instructor for each module; and, finally, the total number of weeks for each semester. Let us assume that:

SAC: Student Attendance's Credibility for each module
A: Average of student attendance for each module
C1: Number of attendance occurrences taken by instructor for each module
C2: Total number of weeks for each semester

Using the above notations, equation 2 can be derived, which measures SAC based on A, C1, and C2. The result will always be always a number between 0 and 1. The purpose of finding SAC of each module is to find which module has a good attendance rate compared to other modules. The values of the average student attendance for each module, the number of attendance occurrences taken by instructor for each module and the total number of weeks for each semester are affecting SAC value.

$$\text{SAC} = \frac{1}{100}\left(A * \frac{c1}{c2}\right) \quad (2)$$

We can simplify the previous equation to be:

$$\text{SAC} = \frac{A\ c1}{100\ c2}$$

$$\frac{A}{100} * \frac{c1}{c2}, \text{ But } A \in [0\text{-}100]$$

Let us assume that Z1 denotes the attendance average divided by 100 (the left hand side of the multiplication)

$$Z1 = \frac{A}{100}$$

$$Z1 \in [0\text{-}1]$$

And, Let us assume that Z2, which is a ratio representing the instructor's overall recording of student attendance throughout the semester, denotes the number of attendance occurrences taken by instructor for each module divided by the total number of weeks for each semester (right hand-side of the multiplication)

$$Z2 = \frac{c1}{c2}$$

$$Z2 \in [0\text{-}1]$$

Therefore, the more attendance occurrences taken by instructor the closer to 1 Z2 will approach.

$$0 \leq Z1\ \&\ Z2 \leq 1$$

The closer to 1 Z1 and Z2 are approaching, the more reliable the student attendance is (i.e. R approaches 1).

By taking the limits of the credibility equation when the number of weeks approaches n, equation (3), and when the number of attendance occurrences taken by instructor for each week approaches the number of weeks per semester (i.e. if the instructor is taking the attendance each time throughout the semester) and, at the same time, when the average value of student attendance is approaching 100% (i.e. when the number of students attending a certain modules is equal the number of registered students at the same module) as in equation (4).

$$(\text{Lim}_{c2 \to \infty} \frac{A\, c1}{100\, c2} = 0) \quad (3)$$

$$(\text{Lim}_{c1 \to c2\, \&\, A \to 100} \frac{A\, c1}{100\, c2} = 1) \quad (4)$$

From equations 3 and 4, it is clear that when the number of weeks increases while fixing the values of A, and C1, the limit of SAC will approach 0, which means that in order to achieve higher credibility (i.e. =1) C1 and C2 must be equal (i.e. instructor must consider taking attendance every week during the semester), and A must be equal to 100 (i.e. number of students attending every week equals to number of students registered at the module).

As an example, by applying equation 2 on three different computer science modules (subjects) - given the codes: 14COC180, 14COA105, and 14COF180 -that represent a sample from the 59 modules' data that were gathered from Computer Science undergraduates of Loughborough University in the UK during the years 2014-2015. We have found that the first module (in this case, 14COC180) has a SAC ratio of 0.067, the second module (14COA105) has a SAC value of 0.349, and the third module (14COF180) has a SAC value of 0.812. We conclude that the first module has the lowest credibility because its value is the smallest and that could be caused by different reasons, such as low number of attendance occurrences taken by the instructor for this module, or low number of students attending the module in a certain week compared to total number of students registered on this module (i.e. low average). The third module has the highest credibility, while the second module has medium credibility.

In order to distinguish between SAC and the occurrence of student's attendance at a certain module, values of SAC at each module should be compared. For example, in module 14COC180, the attendance of the student was 100% knowing that the instructor took the attendance only once during the whole semester. Therefore, the value of 100% will not reflect the real attendance of the student since the module's SAC is very low 0.067. In the second module (14COA105), the attendance of the student was 50% knowing that the instructor took the attendance seven times and that also does not reflect the real attendance because the module's SAC is 0.349. On the other hand, third module (14COF180), the attendance of the student was 70% knowing that the instructor took the attendance eleven times (out of eleven weeks: total number of weeks in this semester). The 70% SAC ratio now reflects the real attendance because the module's SAC is high 0.812 and Z2=1. This draws a conclusion that the higher the SAC, the more accurate the student attendance value will be.

Based on the above, SAC can be seen as an equation that not only takes into account the students' role as an indicator of their attendance, but also the instructor's role: the student's role is represented by A since it measures the student average attendance throughout the semester, while the instructor's role is represented by Z2 which refers to the ratio of taking attendance for the whole semester. That is, a perfect A does not reflect the student's real attendance unless it is complemented with a perfect Z2, and vice versa.

*B. Reliability of SAC*

In order to consider the new measure (SAC) for future uses, its reliability and validity should be assessed. The reliability of a measure refers to the extent to which SAC measurement is consistent [13]. For this purpose, SAC is measured for the same throughout a period of 5 academic semesters from 2010 to 2015 and SAC values are then compared so that Cronbach's Alpha coefficient can be measured to determine the SAC's reliability [14].

10 modules taken by students every year have been chosen randomly amongst other modules. SAC was measured for the modules for each year starting from the academic year 2010/2011 until the 2014/2015, and the results were as shown in Table I below:

TABLE I: SAC FOR 10 RANDOM MODULES OVER 5 YEARS (XX IS A VARIABLE OF THE YEAR)

| Year / Module | 2010/ 2011 | 2011/ 2012 | 2012/ 2013 | 2013/ 2014 | 2014/ 2015 |
|---|---|---|---|---|---|
| XXCOA101 | 0.596 | 0.584 | 0.641 | 0.720 | 0.801 |
| XXCOA122 | 0.542 | 0.681 | 0.636 | 0.574 | 0.552 |
| XXCOB101 | 0.412 | 0.407 | 0.439 | 0.557 | 0.601 |
| XXCOB231 | 0.223 | 0.482 | 0.399 | 0.265 | 0.538 |
| XXCOB232 | 0.335 | 0.389 | 0.417 | 0.416 | 0.505 |
| XXCOB290 | 0.340 | 0.564 | 0.496 | 0.623 | 0.713 |
| XXCOB301 | 0.559 | 0.466 | 0.628 | 0.358 | 0.592 |
| XXCOC003 | 0.389 | 0.423 | 0.371 | 0.444 | 0.556 |
| XXCOC101 | 0.538 | 0.505 | 0.536 | 0.593 | 0.703 |
| XXCOC104 | 0.484 | 0.464 | 0.503 | 0.428 | 0.489 |
| XXCOA101 | 0.596 | 0.585 | 0.641 | 0.72 | 0.801 |

With the aim of obtaining a reliable measure, Cronbach's Alpha internal consistency (Reliability) was applied; although it defines the consistency of the results delivered in a test, ensuring that the various items measuring the different constructs deliver consistent scores [15]. However, in this paper, it was used to calculate the reliability of SAC measure for 10 modules over a period of 5 years.

The following Table II shows the processes applied on the data, resulting an Alpha value of 0.81which ensures high reliability of the SAC.

TABLE II: MEASURING CRONBACH'S ALPHA OF SAC

| | Year10/11 | Year11/12 | Year12/13 | Year13/14 | Year14/15 | TOTAL |
|---|---|---|---|---|---|---|
| XXCOA101 | 0.596 | 0.585 | 0.641 | 0.720 | 0.801 | 3.343 |
| XXCOA122 | 0.543 | 0.681 | 0.636 | 0.574 | 0.552 | 2.986 |
| XXCOB101 | 0.412 | 0.407 | 0.439 | 0.557 | 0.601 | 2.416 |
| XXCOB231 | 0.223 | 0.482 | 0.399 | 0.265 | 0.538 | 1.906 |
| XXCOB232 | 0.335 | 0.389 | 0.417 | 0.416 | 0.505 | 2.063 |
| XXCOB290 | 0.340 | 0.564 | 0.496 | 0.623 | 0.713 | 2.735 |
| XXCOB301 | 0.559 | 0.466 | 0.628 | 0.358 | 0.592 | 2.604 |
| XXCOC003 | 0.389 | 0.423 | 0.371 | 0.444 | 0.556 | 2.183 |
| XXCOC101 | 0.538 | 0.505 | 0.536 | 0.593 | 0.703 | 2.875 |
| XXCOC104 | 0.484 | 0.464 | 0.503 | 0.428 | 0.489 | 2.368 |
| Total | 4.419 | 4.966 | 5.067 | 4.978 | 6.050 | 25.481 |
| Var | 0.015 | 0.008 | 0.010 | 0.019 | 0.010 | 0.063 |

| | |
|---|---|
| k | 5 |
| ∑Var | 0.063 |
| var | 0.180 |
| α | 0.815 |

where:

k: is the number of years

Var: is the variance of SAC values for each module over one year only

∑Var: is the sum of Var values over the period of 5 years

var: is the population variance, and

α: is the Cronbach's Alpha whose values can be between 0 and 1.0 (the higher is the more reliable) and its equation can be given as:

$$(\frac{k}{k-1}) * (1 - \frac{\sum \text{var}}{\text{var}}) \quad \ldots \text{Cronbach's Alpha [16]}.$$

Based on the above results, it can be concluded that the newly proposed measure (SAC) can be relied on when dealing with student attendance data; hence it can be considered as a reliable measure.

## IV. CLASSIFYING SAC

The aim of classifying SAC is to identify the attributes that can affect the final SAC values. In other words, the classification process will highlight the attributes that significantly affect the module's SAC.

In this paper, data set was analyzed using J48 decision tree classification technique. The J48 DT technique is characterized by the ease of rules generation and ease of understanding. The J48 DT works as follows: In order to classify a new data item, firstly, a DT has to be created based on the attribute values of the available training data (i.e. 70% out of the available 59 modules' data). It uses the fact that each attribute of the data can be used to make a decision by splitting the data into smaller subsets. J48 examines the normalized information gain that results from choosing an attribute for splitting the data. The J48 makes the decision by using the attribute with the highest normalized information gain. Then the algorithm recurs on the smaller subsets [17]. In this paper, the J48 recursively split the tree based on choosing the Average Student Attendance as the attribute with the highest normalized gain. In fact, for each iteration, the J48 examines the information gain for all the attributes again, and then chooses the one with the highest information gain as follows:

The information gain can be given as:
where *p* is the probability of the attribute [18].

Hence, by examining any data record (i.e. set of attributes) against the above formula, the J48 ranks the independent attributes based on their information gain as shown in Table III below:

TABLE III: SELECTED ATTRIBUTES AND THEIR INFORMATION GAIN

| Attribute | Information Gain |
|---|---|
| attend_avg | 2.357 |
| reg_modulecode | 2.352 |
| attend_taken | 0.728 |
| reg_semester | 0.138 |

Therefore, it can be clearly shown that the J48 has chosen the Average Student Attendance for its relatively high information gain.

In order to apply this technique following steps are performed in sequence:

### A. Data Selection and Transformation

In this step, only the attributes that are required for processing with data mining were selected. The selection was based on the value those attributes add. In other words, some attributes add no value to the measurement per se, like the Module Date or Attendance Reason, hence they were neglected since it is not crucial to determine the average student attendance, for example, for attendance occurrences that have certain reason, all occurrences count. On the contrary, other attributes represent the base of the measurement, such as the Attendance Status. All related variables derived from the database are given in Table IV.

The table shows the selected variables and their possible values; the attend_avg denotes the Average of Student Attendance, whose value can be calculated using equation 1, attend_taken represent the number of occurrences the instructor has taken the attendance. Its values range from 1 to 11 times, where 11 is the maximum number of weeks during which the instructor may take attendance. SAC is a value that can be calculated using equation 2, and its values range from 0-1. SAC_Strength is derived directly from the SAC values in a way J48 can handle the SAC as nominal values. That is, in order to overcome one of the main drawbacks of using J48, which is the fact that it does not classify numerical values, SAC (i.e. numerical) had to be represented as nominal values in order to be classified properly by the J48 algorithm. Table V below shows the nominal representation of SAC.

TABLE IV: STUDENT RELATED VARIABLES

| Variable | Description | Possible Values |
|---|---|---|
| Module_code | Module code | All Modules codes |
| attend_avg | Average of Student Attendance | 0 – 100 (%) |
| attend_taken | Number of student attendance occurrences taken by instructor | 1-11(times) |
| SAC | Student Attendance's Credibility | 0.0 – 1.0 |
| sem_no | Semester number | 1, 2 |
| SAC_Strength | Strength of Credibility (Nominal) | 1- 10 |

TABLE V: NOMINAL REPRESENTATION OF CREDIBILITY

| SAC (Numeric) | Strength of SAC (Nominal) |
|---|---|
| [0 – 0.1) | 1 |
| [0.1 – 0.2) | 2 |
| [0.2 - 0.3) | 3 |
| [0.3 – 0.4) | 4 |
| [0.4 – 0.5) | 5 |
| [0.5 – 0.6) | 6 |
| [0.6 – 0.7) | 7 |
| [0.7 – 0.8) | 8 |
| [0.8 – 0.9) | 9 |
| [0.9 – 1.0] | 10 |

### B. Classification Process

Table IV shows a number of variables through which the J48 built the decision tree. All variables, except for SAC are independent variables, whereas SAC is a dependent variable.

As mentioned before, the J48 will start by examining the normalized information gain for each of the independent variables and by doing so, it recursively splits the tree into branches. As an example, the J48 starts examining the information gain of each of the attributes, starting with Average Attendance.

## V. RESULTS AND DISCUSSION

The results of the training test using J48 Decision Tree showing that when the output is the SAC_Strength; the correctly classified instances (Accuracy) is 88.9 %, the incorrectly classified instances is 11.1 %, and the Root mean squared error is 0.1337.

Figure 2 shows the tree diagram for the training set when the output is the SAC_Strength. As mentioned earlier, 59

modules were used, 70% of which are split for training. The number of leaves in the tree is 6 and the size of the tree is 11 and the most significant factor is attendance average.

According to the results shown in Fig. 2, the rules presented in Table VI will be generated.

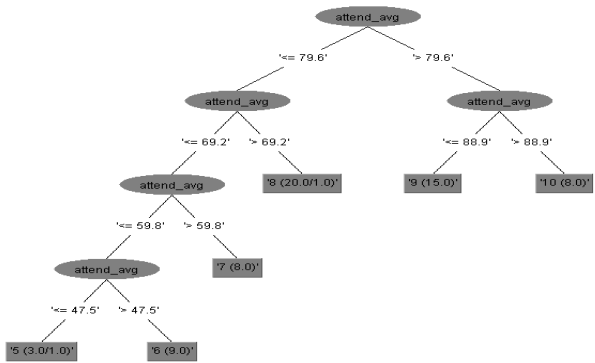

Fig. 2. Tree diagram for the training set.

TABLE VI: GENERATED RULES OF THE TRAINING SET

| Rule No. | Generated Rules |
|---|---|
| 1 | If attend. avg>88.9 then: SAC_Strength = 10 |
| 2 | If attend. avg<= 88.9 and attend. avg>79.6 then: SAC_Strength = 9 |
| 3 | If attend. avg<79.6 and attend. avg>= 69.2 then: SAC_Strength = 8 |
| 4 | If attend. avg<69.2 and attend. avg>= 59.8 then: SAC_Strength = 7 |
| 5 | If attend. avg<59.8 and attend. avg>= 47.5 then: SAC_Strength = 6 |
| 6 | If attend. avg=<47.5 then: SAC_Strength =5 |

As shown in Table VI, it is clear that the average value of student attendance represents a significant factor in determining the value of SAC. For example, rule number 1 shows that whenever the average rises above 88.9%, SAC is classified as 10 (i.e. strong credibility). On the contrary, when the average value of the student attendances is equal or below 47.5%, SAC is classified as 5 (i.e. poor credibility).

Not surprisingly, although all selected attributes contribute to the credibility of student attendance, however, it is clear that the most significant factor on SAC is the average of their attendance throughout the semester, as shown in section 4. That is, the sum of the ratios between the number of students attending a module to the number of students registered to that module over the period of 11 weeks.

## VI. CONCLUSIONS AND FUTURE WORK

The main objective of this work has been achieved by measuring the credibility of student attendance for each module depending on two main factors: the average of student attendance and the number of attendance occurrences taken by the instructor during the semester.

Student Attendance's Credibility (SAC) for each module helps in deciding whether an attendance record about certain module is reliable or not based not only on number of attendance occurrences taken by a module instructor, but also by various significant factors such as the average value of student attendance, and the number of students attending a module each week compared to the number of students registered at the module.

The results have shown that even if a student's attendance was 100% during a semester, this value does not necessarily reflect the real attendance of the student unless it is compared to the SAC of the module itself and to the Z2 value in particular, and that is due to other factors that were taken into consideration in this study.

J48 DT data mining classification technique was utilized to classify SAC based on the strength classes proposed in this study, as shown in table 6. This has resulted in classifying SAC into different strength classes rather than considering SAC as just a number between 0 and 1. Based on that, classification rules were also generated based on the average value of the student attendance which represented the most significant factor on measuring SAC.

As an extension of this paper, the transcripts data and the graduation survey of a certain student along with his/her attendance's SAC achieved by this work will be taken into consideration to build a predication model that assist the student in making a decision regarding his/her future career.

**Mohammed Alsuwaiket** got the bachelor degree from University of Bradford, UK, 2006, in information technology management; the master degree from University of Hertfordshire, UK, 2012, in computer science; and the PhD student, University of Loughborough, UK, in computer science.

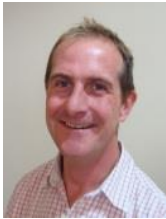

**D. Christian** is with the Loughborough University, Epinal Way, Loughborough Leicestershire, UK. LE11 3TU.

He is a senior lecturer in the Department of Computer Science at Loughborough University. He joined the Department in October 1999 from the University of Derby (1993 - 1999) where he began as a lecturer and moved on to become a School Reader. His research interests are primarily in the application and development of artificial neural networks to rainfall-runoff modelling and flood forecasting (neurohydrology) although he also involved in research and consultancy in software project management.

Dr. Dawson is an editorial board member of Computational Intelligence and Neuroscience, External Examiner - Nottinngham Trent University, UK 0 undergraduate degree programmes in Computer Science, DMIN Programme Committee Member - 10th International Conference on Data Mining, Las Vegas, 21-24 July, 2014., Elected member of the IACCS Board (International Association of Computer Science in Sport)., and DMIN Programme Committee Member - 9th International Conference on Data Mining, Las Vegas, 22-25 July, 2013.

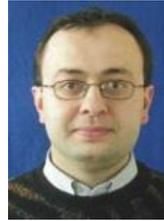

**B. Firat** delivered the computer science BSc degree foundation subjects at London Meridian College between 1997 and 2001. He received a BSc degree from Computer Engineering Department of Ege University, Izmir, Turkey in 1997. He obtained a database systems MSc degree from University of Westminster in 2002 and a PhD degree from the Department of Computer Science, Loughborough University in 2011.

He has been a university teacher at Department of Computer Science, Loughborough University since 2007. He was a visiting lecturer at School of Electronics and Computer Science, University of Westminster, London before he is employed by Loughborough University as a teaching assistant in 2003.

Dr. Batmaz is co-ordinator of Erasmus, Widening Parcitipation, Part A Tutor, and Member of the Software Based Systems (SBS) interest group.